\documentclass[eqsecnum,preprintnumbers,showpacs,aps]{revtex4}

\usepackage{color}

\usepackage{graphicx}% Include figure files
\usepackage{amsmath,amssymb}
\usepackage{bm}

\newcommand{\beq}{\begin{equation}}
\newcommand{\eeq}{\end{equation}}

\def \({\left( }
\def \){\right) }
\newcommand{\beqar}{\begin{eqnarray}}
\newcommand{\eeqar}{\end{eqnarray}}

\newcommand{\gsim}{\, \raisebox{-0.8ex}{$\stackrel{\textstyle >}{\sim}$ }}
\newcommand{\lsim}{\, \, \raisebox{-0.8ex}{$\stackrel{\textstyle <}{\sim}$ }}

\newcommand{\dalm}{\kern1pt\vbox{\hrule height 0.9pt\hbox{\vrule width 0.9pt\hskip 2.5pt\vbox{\vskip 5.5pt}\hskip 3pt\vrule width 0.3pt}\hrule height 0.3pt}\kern1pt}

\begin{document}
\preprint{YITP-14-15}

%\draft
%@twocolumnfalse\endcsname

% For two column
%\wideabs{

\title{Scalar gravitational waves from relativistic stars in scalar-tensor gravity}

\author{Hajime Sotani}
\affiliation{
Yukawa Institute for Theoretical Physics, Kyoto University, Kyoto 606-8502, Japan
}

\date{\today}

% Abstract
\begin{abstract}
Unlike general relativity, the scalar gravitational waves can be excited due to the radial oscillations in scalar-tensor gravity. To examine the scalar gravitational waves in scalar-tensor gravity, we derive the evolution equations of the radial oscillations of neutron stars and determine the specific oscillation frequencies of the matter oscillations and scalar gravitational waves, where we adopt two different numerical approaches, i.e., the mode analysis and direct time evolution. As a result, we observe the spontaneous scalarization even in the radial oscillations. Depending on the background scalar field and coupling constant, the total energy radiated by the scalar gravitational waves dramatically changes, where the specific oscillation frequencies are completely same as the matter oscillations. That is, via the direct observations of scalar gravitational waves, one can not only reveal the gravitational theory, but also extract the radial oscillations of neutron stars. 
\end{abstract}

%
%%%%%%%%%%%%%%%%%%%%%%%%%%%%%%%%%%%%%%%%%%%%%%%%%%%%%%%%%%%%%%%
%  04.50.Kd :  Modified Theories of Gravity
%  04.80.Cc :  Experimental tests of gravitational theories
%  04.40.Dg :  Relativistic stars: structure, stability, and oscillations (see also 97.60.-s Late stages of stellar evolution) 
%  97.60.Jd :  Neutron stars (see also 26.60.+c Nuclear matter aspects of neutron stars in nuclear physics) 
%%%%%%%%%%%%%%%%%%%%%%%%%%%%%%%%%%%%%%%%%%%%%%%%%%%%%%%%%%%%%%%

%]

% For two column
%}

\maketitle

%\baselineskip 24pt

%%%%%%%%%%%%%%%%%%%%%%%%%%%%%%%%%%%%%%%%%%%%%%%%
\section{Introduction}
\label{sec:I}
%%%%%%%%%%%%%%%%%%%%%%%%%%%%%%%%%%%%%%%%%%%%%%%%

Since general relativity has been proposed, many experiments have been performed to verify the gravitational theory. Most of these attempts are done in the weak gravitational field such as our solar system, but nothing indicates the failure of general relativity. On the other hand, since the astronomical observations in the strong gravitational field are very poor, the gravitational theory in such a strong-field regime could be still unconstrained. That is, the gravitational theory to describe the phenomena in the strong-field regime might be different from general relativity, and one might be possible to probe the gravitational theory via the observations of the deviation from general relativity. In practice, up to now there are many suggestions to observationally test the gravitational theory in the strong-field regime \cite{W1993,W2001,P2008}. The technology is developing more and more, which will enable us to accurately observe the phenomena in the strong-field regime. These coming new observations might be possible to use as the test of gravitational theory.

So far, a lot of alternative gravitational theories are proposed. Among them, scalar-tensor theory is the one of the simplest alternative gravitational theories, which must be a natural extension of standard general relativity \cite{DE1992}. One of the motivations to consider scalar-tensor gravity is that this theory can be obtained in the low energy limit of string and/or other gauge theories. In scalar-tensor gravity, the scalar field plays an essential role in addition to the usual tensor field in general relativity, where the matter field is described by using the effective metric $\tilde{g}_{\mu\nu}$ associated with the scalar and gravitational field, $\varphi$ and $g_{*\mu\nu}$, via the conformal transformation, i.e., $\tilde{g}_{\mu\nu}=A^2(\varphi)g_{*\mu\nu}$. The Brans-Dicke theory \cite{BD} is the simplest version of scalar-tensor gravity, where $A(\varphi)$ is defined as $A(\varphi)=exp(\alpha\varphi)$. The coupling parameter $\alpha$ can be associated with the Brans-Dicke parameter $\omega_{\rm BD}$ as $\alpha^2=1/(2\omega_{\rm BD}+3)$, which are constrained through the solar system experiments, i.e., $\omega_{\rm BD} \gsim 40000$ which is corresponding to $\alpha < 10^{-5}$ \cite{EF2004}. Within this restriction on the coupling parameter, it is almost impossible to predict a large deviation from general relativity in the strong-field regime.

A different functional form of the conformal factor is also suggested by Damour and Esposito-Far\`{e}se \cite{DE1993,DE1996}, where $A(\varphi)\equiv exp(\alpha\varphi + \beta\varphi^2/2)$. With this type of coupling, even if $\alpha$ is almost zero, the relativistic stellar models in scalar-tensor gravity can significantly deviate from the predictions in general relativity. Additionally, they found that the stellar models in scalar-tensor gravity suddenly deviate from those in general relativity for the specific values of coupling parameters, which is referred to as {\it spontaneous scalarization}. With respect to this phenomenon, Harada systematically examined with the technique of catastrophe theory and found that the spontaneous scalarization can happen for $\beta\lsim -4.35$ \cite{H1998}. Recently, it is found that the spontaneous scalarization are possible for larger value of $\beta$ in fast rotating relativistic stars \cite{Daniela2013} and in the neutron star binary system \cite{BPPL2013,Shibata2013,PBPL2013}. On the other hand, using the observations of pulsar white dwarf binary, Freire {\it et al}. set a severe constraint on $\beta$, i.e., $\beta\gsim -5$ \cite{F2012}. 
Additionally, it is reported that $\beta$ could be constrained to be larger than $-4.5$, depending on the equation of state \cite{Shibata2013}. 
Maybe, although the constraint on $\beta$ would become severer via the future observations, we focus on the range of $\beta \gsim -5$ in this paper.

The several attempts to observationally distinguish scalar-tensor gravity from general relativity have been already done in the past by using the redshift in the absorption lines of the X and $\gamma$ rays emitted from the stellar surface \cite{DP2003}, the spectrum of the gravitational waves radiated from relativistic stars \cite{SK2004,SK2005}, and the rotational effect around compact objects \cite{Sotani2012}. In this paper, we consider the different approach, i.e., the scalar gravitational waves driven by the radial oscillations. In fact, the gravitational waves can not be excited due to the radial oscillations in general relativity. This means that the detection of scalar gravitational waves itself becomes the proof the existence of scalar field. From the observational point of view,
if the scalar gravitational waves exist, one could in principle identify the scalar gravitational waves with more than three gravitational wave detectors, 
because we have only three degrees of polarizations in scalar-tensor gravity, such as two usual tensor gravitational waves and scalar gravitational wave. In fact, we expect that five gravitational wave detectors will be in operation in the future, such as two advanced LIGOs \cite{aLIGO}, advanced Virgo \cite{aVirgo}, KAGRA \cite{KAGRA}, and IndIGO \cite{IndIGO}. On the other hand, the method how to separate and reconstruct an arbitrary number of polarization modes by using the observational data by multiple interferometric gravitational wave detectors, is also developing, which is a model-independent approach \cite{HN2013}. 
It should be notices that the scalar gravitational waves in physical frame are proportional to the cosmological value of scalar field \cite{BPPL2013,Shibata2013}, whose value must be quite small. That is, if the scalar gravitational waves exist, they might be quite weak and difficult to detect in the current gravitational wave detectors.

The radial oscillations of relativistic stars in general relativity have been examined since early times \cite{C1964,GL1983,GL1992,KR2001} in the context of the stability analysis. Meanwhile, the scalar gravitational waves in scalar-tensor gravity are also examined in the black hole formation due to the dust collapse \cite{Shibata1994,Scheel1995,H1997} and the test particle around a Kerr black hole \cite{Saijo1997}. Anyway, this is the first time to calculate the scalar gravitational waves driven by the stellar radial oscillations in scalar-tensor gravity suggested by Damour and Esposito-Far\`{e}se \cite{DE1993,DE1996}. For this purpose, we will derive the perturbation equations of radial oscillations and make numerical calculations to examine it.

This paper is organized as follows. In the next section, we briefly mention the equilibrium of nonrotating relativistic stars in scalar-tensor gravity. In Sec. \ref{sec:III}, we derive the perturbation equations describing the radial oscillations of relativistic stars in scalar-tensor gravity. The numerical results are shown in Sec. \ref{sec:IV}, where the specific frequencies are determined with the mode analysis and the direct time evolution. Then, we make a conclusion in Sec. \ref{sec:V}. We adopt the geometric units, $c=G_*=1$, where $c$ and $G_*$ denote the speed of light and the gravitational constant, respectively, and use the metric signature is $(-,+,+,+)$.

%%%%%%%%%%%%%%%%%%%%%%%%%%%%%%%%%%%%%%%%%%%%%%%%
\section{Stellar models in Scalar-Tensor gravity}
\label{sec:II}
%%%%%%%%%%%%%%%%%%%%%%%%%%%%%%%%%%%%%%%%%%%%%%%%

In particular, in this paper, we consider the neutron star models in scalar-tensor theory of gravity with one scalar field. In fact, this is a natural extension of general relativity, where gravity is mediated not only by a usual tensor field, but also by a massless long-range scalar field. To express such a theory, the total action in the Einstein frame is given by Ref. \cite{DE1992}
\begin{equation}
  S = \frac{1}{16\pi G_*}\int\sqrt{-g_*}\left(R_*-2g_*^{\mu\nu}\varphi_{,\mu}\varphi_{,\nu}\right)d^4x
      + S_m\left[\Psi_m,A^2(\varphi)g_{*\mu\nu}\right],
\end{equation}
where $G_*$ is the bare gravitational constant, $R_*$ is the scalar curvature determined by the Einstein metric $g_{*\mu\nu}$, $\varphi$ is the scalar field, and $\Psi_m$ represents collectively all matter fields. The metric tensor in the Einstein frame, $g_{*\mu\nu}$, is related to that in the physical frame (or Jordan-Fierz frame), $\tilde{g}_{\mu\nu}$, as
\begin{equation}
  \tilde{g}_{\mu\nu} = A^2(\varphi) g_{*\mu\nu}.
\end{equation}
Hereafter, in order to clarify the frame, the quantities in the physical frame are denoted by a tilde and those in the Einstein frame are denoted by an asterisk. We remark that the field equations are usually formulated in the Einstein frame, but all non-gravitational experiments are observed in the physical frame.

Varying the total action $S$, one can get the field equations in the Einstein frame for the tensor and scalar fields;
\begin{gather}
  G_{*\mu\nu} = 8\pi G_* T_{*\mu\nu} +  T^{(\varphi)}_{*\mu\nu}, \label{eq:field1} \\
   \dalm_* \varphi = -4\pi G_* \alpha(\varphi)T_*, \label{eq:field2}
\end{gather}
where $T_{*\mu\nu}$ is the energy-momentum tensor of the fluid in the Einstein frame, while $T^{(\varphi)}_{*\mu\nu}$ denotes the energy-momentum tensor of the massless scalar field, i.e.,
\begin{equation}
  T^{(\varphi)}_{*\mu\nu} \equiv 2\varphi_{,\mu}\varphi_{,\nu} 
      - g_{*\mu\nu}g_*^{\alpha\beta}\varphi_{,\alpha}\varphi_{,\beta}.
\end{equation}
$T_{*\mu\nu}$ is associated with the energy-momentum tensor in physical frame $\tilde{T}_{\mu\nu}$ as
\begin{equation}
  T_*^{\mu\nu} \equiv \frac{2}{\sqrt{-g_*}}\frac{\delta S_m}{\delta g_{*\mu\nu}} = A^6(\varphi)\tilde{T}^{\mu\nu}.
\end{equation}
In Eq. (\ref{eq:field2}), the scalar quantities $T_*$ and $\alpha(\varphi)$ are defined as $T_*\equiv T_*^{\mu\nu}g_{*\mu\nu}$ and $\alpha(\varphi)\equiv d\ln A(\varphi)/d\varphi$. Since $\alpha(\varphi)$ obviously relates a scalar field to matter, the theory with $\alpha(\varphi)=0$ exactly reduces to general relativity. In addition to the field equations, the law of energy-momentum conservation is given as $\tilde{\nabla}_\nu\tilde{T}_{\mu}^{\ \nu}=0$ in the physical frame, which is transformed into that in the Einstein frame, such as
\begin{equation}
  \nabla_{*\nu}T_{*\mu}^{\ \ \nu}=\alpha(\varphi)T_*\nabla_{*\mu}\varphi. \label{eq:field3}
\end{equation}
In this paper, we adopt the same form of conformal factor $A(\varphi)$ as in Damour \& Esposito-Far\`{e}se \cite{DE1993}, i.e.,
$A(\varphi) = \exp\left(\beta\varphi^2/2\right)$,
where $\beta$ is a real number. With this conformal factor, the quantity $\alpha(\varphi)$ is expressed as $\alpha(\varphi)=\beta\varphi$, i.e., the theory with $\beta=0$ agrees with general relativity. At last, we set $\varphi_0$ as the cosmological value of scalar field at infinity. In particular, we adopt $\varphi_0=0$ in this paper.

\begin{figure*}
\begin{center}
\begin{tabular}{cc}
\includegraphics[scale=0.5]{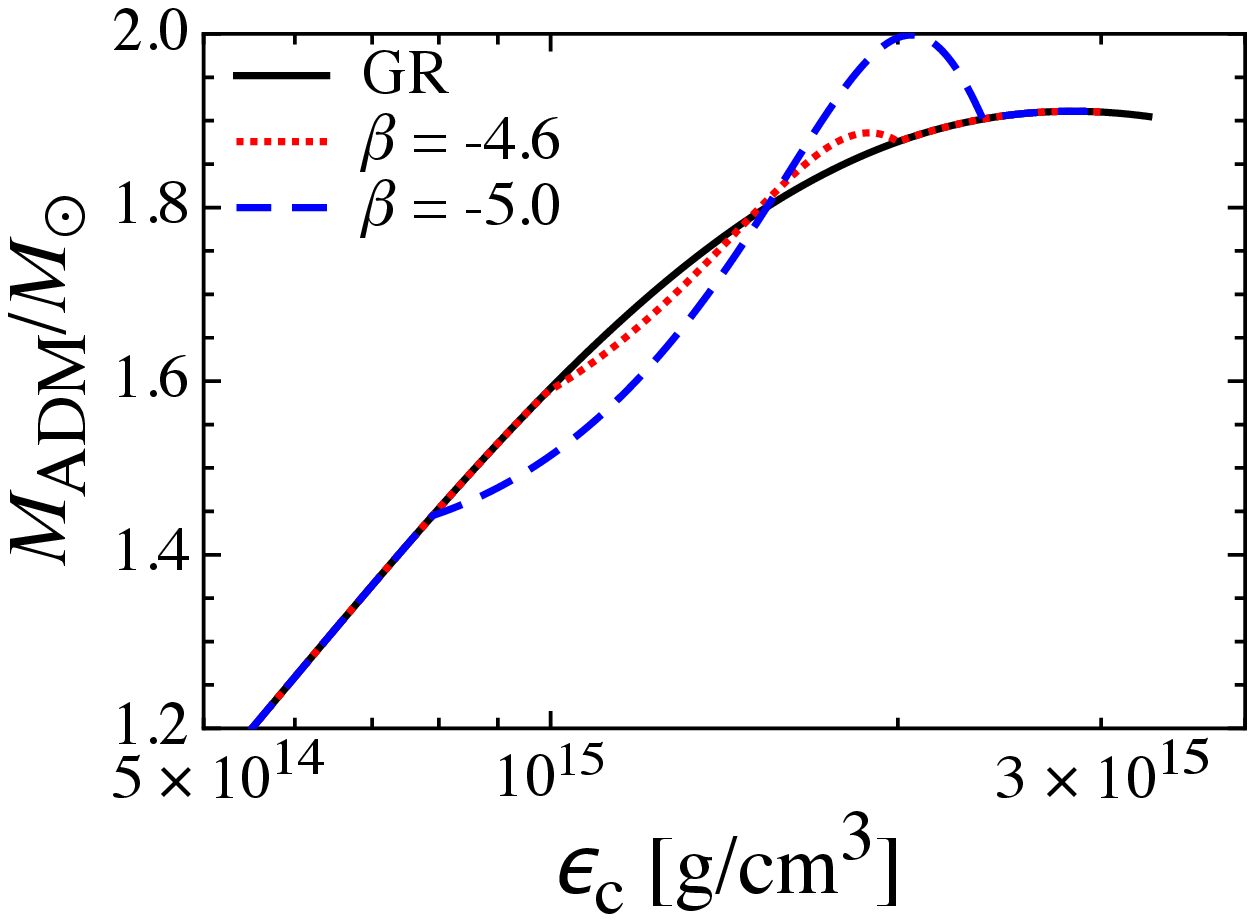} &
\includegraphics[scale=0.5]{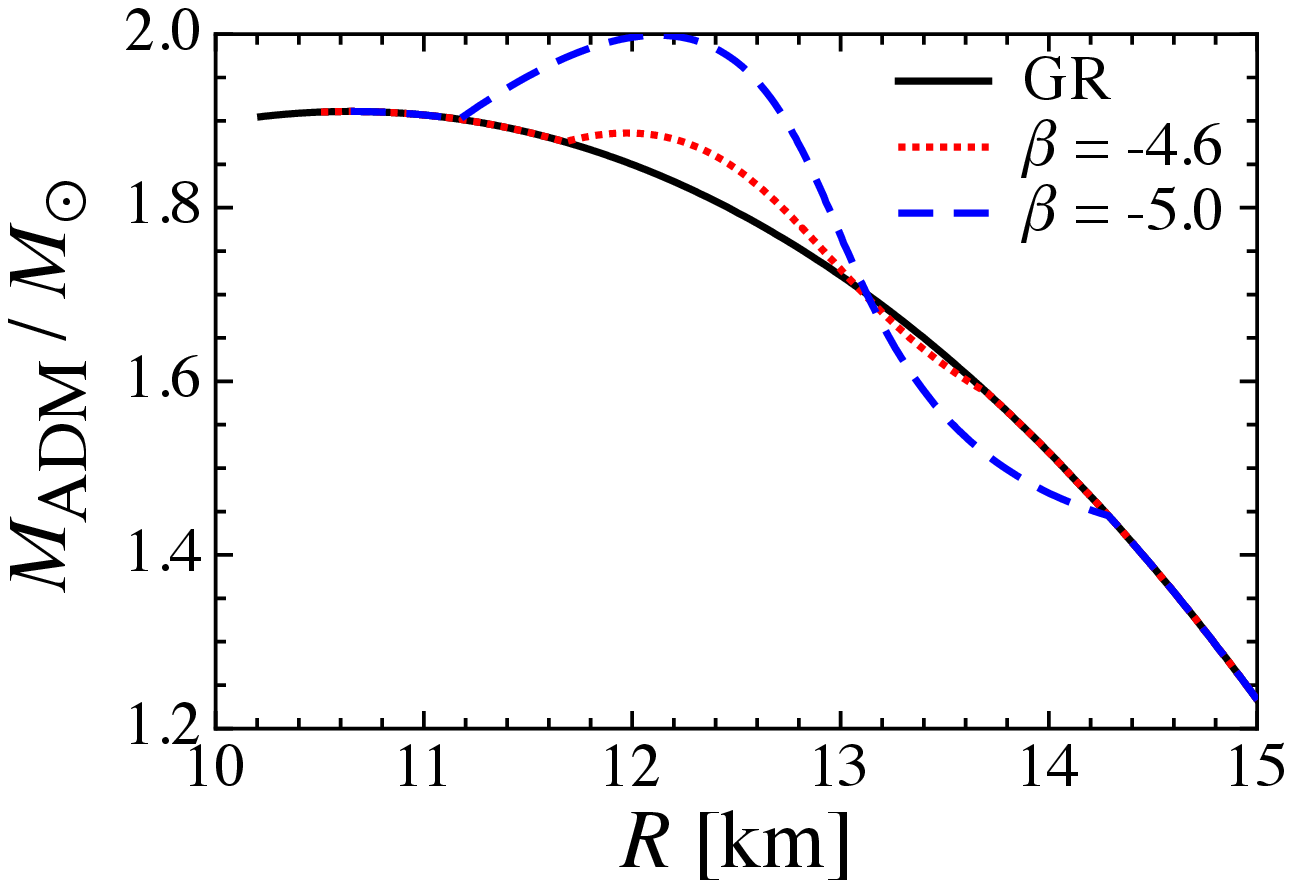}
\end{tabular}
\end{center}
\caption{%%
(Color online) Neutron star models in scalar-tensor gravity with $\beta=-5.0$ (broken lines) and $\beta=-4.6$ (dotted lines), and in general relativity with the solid lines. The left and right panels are corresponding to the ADM masses of neutron stars as functions of the central density and stellar radius, respectively.
}%%
\label{fig:NS}
\end{figure*}

\begin{figure*}
\begin{center}
\includegraphics[scale=0.5]{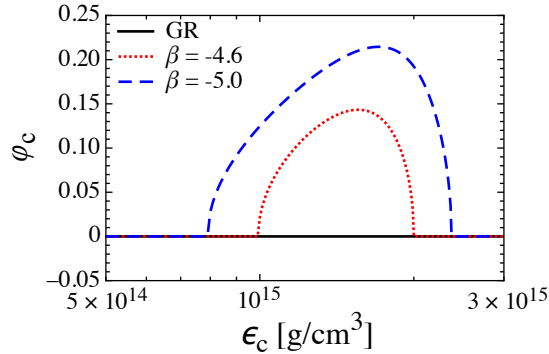} 
\end{center}
\caption{%%
(Color online) Central values of scalar field, $\varphi_c$, as a function of the central density, where the broken and dotted lines correspond to the neutron star models in scalar-tensor gravity with $\beta=-5.0$ and $-4.6$ while the solid line corresponds to that in general relativity.
}%%
\label{fig:phic}
\end{figure*}

Now, we consider the neutron star models in scalar-tensor theory, which are constructed with perfect fluid of cold degenerate matter. The metric for non-rotating, spherically symmetric neutron star models can be described as
\begin{align}
  ds_*^2 & = g_{*\mu\nu}dx^\mu dx^\nu \nonumber \\
               & =  -e^{2\Phi}dt^2 + e^{2\Lambda}dr^2 + r^2 (d\theta^2 + \sin^2\theta d\phi^2),
\end{align}
where $\Phi$ and $\Lambda$ are functions of $r$, and $e^{2\Lambda}$ is associated with the mass function $\mu(r)$ as $e^{-2\Lambda}=1-2\mu(r)/r$. With respect to the matter,  we assume the perfect fluid;
\begin{equation}
  \tilde{T}_{\mu\nu} = \left(\tilde{p} + \tilde{\epsilon}\right)\tilde{u}_{\mu}\tilde{u}_\nu + \tilde{p}\tilde{g}_{\mu\nu},
\end{equation}
where $\tilde{u}_\mu$, $\tilde{p}$, and $\tilde{\epsilon}$ are the four-velocity of fluid, pressure, and total energy density in physical frame.  In particular, the four-velocity of equilibrium neutron star models is given by 
\begin{equation}
  \tilde{u}^\mu = (A^{-1}e^{-\Phi},0,0,0).
\end{equation}
Then, the equilibrium models are determined by integrating the Tolman-Oppenheimer-Volkoff (TOV) equations in scalar-tensor theory \cite{DE1993,H1998,SK2004}, assuming the relation between the pressure and energy density, i.e., EOS. In this paper, we adopt the polytrope EOS, $\tilde{p}=K\tilde{\epsilon}^{\Gamma}$, where we especially fix that $K=200$ km$^2$ and $\Gamma=2$. In Fig. \ref{fig:NS}, we show the neutron star models in scalar-tensor gravity with $\beta=-5.0$ (broken line) and $\beta=-4.6$ (dotted line) together with the case in general relativity (solid line), where the left and right panels correspond to the Arnowitt-Deser-Misner (ADM) masses of neutron stars as functions of central density $\tilde{\epsilon}_c$ and stellar radius $R$, respectively. Additionally, the central values of scalar field are shown in Fig. \ref{fig:phic} as a function of the central density.
We remark that, as mentioned before, the realistic value of $\beta$ might be larger than $-4.5$ and one could have a chance to observe the scalarization even with $\beta=-4.2$ in the rapidly rotating neutron stars and/or in neutron star binaries \cite{Daniela2013,BPPL2013,Shibata2013,PBPL2013}. But, in the case of spherically symmetric stars, one can observe the scalarization only for $\beta \lsim -4.35$ \cite{H1998}, i.e., the stellar models with for example $\beta=-4.0$ or $-4.2$ are completely equivalent to those in general relativity, where no scalar perturbation is induced by the matter motion. So, in this paper, we especially consider the stellar models with $\beta=-4.6$ and $-5.0$ to examine the stellar oscillations and induced scalar perturbations.

%%%%%%%%%%%%%%%%%%%%%%%%%%%%%%%%%%%%%%%%%%%%%%%%
%  FIGURE 1
%%%%%%%%%%%%%%%%%%%%%%%%%%%%%%%%%%%%%%%%%%%%%%%%
%
%

%%%%%%%%%%%%%%%%%%%%%%%%%%%%%%%%%%%%%%%%%%%%%%%%
%  FIGURE 2
%%%%%%%%%%%%%%%%%%%%%%%%%%%%%%%%%%%%%%%%%%%%%%%%
%
%

%%%%%%%%%%%%%%%%%%%%%%%%%%%%%%%%%%%%%%%%%%%%%%%%
\section{Radial oscillations}
\label{sec:III}
%%%%%%%%%%%%%%%%%%%%%%%%%%%%%%%%%%%%%%%%%%%%%%%%

In this section, we consider the radial oscillations on the stellar models mentioned in the previous section, adopting the relativistic Cowling approximation. The scalar field is described as
\begin{gather}
%  \tilde{g}_{\mu\nu} = \tilde{g}^{\rm (B)}_{\mu\nu} + \tilde{h}_{\mu\nu}, \\
  \varphi = \varphi^{\rm (B)} + \delta \varphi(t,r),
\end{gather}
where $\varphi^{\rm (B)}$ denotes the non-perturbed scalar field, while $\delta \varphi$ corresponds to the perturbations of scalar field. With the Cowling approximation, the metric perturbations in the physical frame are neglected, i.e., $\delta\tilde{h}_{\mu\nu}=0$. However, the metric perturbations in the Einstein frame, $h_{*\mu\nu}$, can be induced by the perturbation of scalar field even with the Cowling approximation, which are given by $h_{*\mu\nu}=-2g_{*\mu\nu}\delta A/A$. So, the metric perturbations in the Einstein frame are expressed as
\begin{gather}
  h_{*tt} = 2\beta\varphi e^{2\Phi}\delta\varphi,  \label{eq:h*tt} \\
  h_{*rr} = - 2\beta\varphi e^{2\Lambda}\delta\varphi, \\
  h_{*\theta\theta} = -2\beta\varphi r^2 \delta\varphi, \\
  h_{*\phi\phi} = -2\beta\varphi r^2\sin^2\theta \delta\varphi, \label{eq:h*phiphi} 
\end{gather}
and the other components are zero, where we use $\delta A/A=\beta\varphi \delta\varphi(t,r)$.
We remark that the Cowling approximation in general relativity is a well defined approach where only fluid dynamics is allowed. But, in scalar-tensor gravity, the fluid motion excites the variation of scalar fields via Eq. (\ref{eq:field2}). So, to examine how the scalar fields are affected by the fluid motion, we consider in this paper that the variations of matter and scalar fields with fixing the physical metric. That is, we assume that the variations of the matter and scalar fields affect little the physical metric. Of course, one can examine only fluid dynamics with fixing the scalar field and physical metric as in Ref. \cite{SK2004}, which may be corresponding to the standard Cowling approximation. On the other hand, we should consider the full linearized problem including the metric perturbations in the future as in Ref. \cite{SK2005}.
%Anyway, we consider that the treatment in this paper could be better than the standard Cowling approximation but be inferior as compared with the full problem.

On the other hand, the fluid perturbations are described by the Lagrangian displacement vector;
\begin{equation}
  \tilde{\xi}^i = (W,0,0),
\end{equation}
where $W$ is a function of $t$ and $r$. Then, the perturbed four-velocity, $\delta \tilde{u}^\mu$, has the form
\begin{equation}
  \delta\tilde{u}^\mu = \frac{1}{Ae^{\Phi}}\left(0, \frac{\partial W}{\partial t},0,0\right).
\end{equation}
At last, the pressure and energy density perturbations are described as
\begin{gather}
  \tilde{p} = \tilde{p}^{\rm (B)} + \delta \tilde{p}(t,r), \\
  \tilde{\epsilon} = \tilde{\epsilon}^{\rm (B)} + \delta \tilde{\epsilon}(t,r),
\end{gather}
where $\tilde{p}^{\rm (B)}$ and $\tilde{\epsilon}^{\rm (B)}$ denote the pressure and energy density in the equilibrium stellar models, while $\delta\tilde{p}$  and $\delta \tilde{\epsilon}$ denote the pressure and energy density perturbations, respectively.

With the above perturbation variables, the perturbation equations are derived by taking the variation of Eqs. (\ref{eq:field2}) and (\ref{eq:field3});
\begin{gather}
%  \delta G_{*\mu\nu} = 8\pi G_*\delta T_{*\mu\nu} + \delta T^{(\varphi)}_{*\mu\nu}, \label{eq:dEin} \\
  \delta (\dalm_* \varphi) = -4\pi G_* \delta\left[\alpha(\varphi)T_*\right], \label{eq:dphi0} \\
  \delta (\nabla_{*\nu}T_{*\mu}^{\ \ \nu}) = \beta \delta\left[\varphi T_*\nabla_{*\mu}\varphi\right]. \label{eq:conservation}
\end{gather}
From Eq. (\ref{eq:dphi0}), one can get the evolution equation with respect to the scalar field;
\begin{align}
  -e^{-2\Phi+2\Lambda}\delta\ddot{\varphi} +& \delta\varphi'' + \left(\eta - 2\beta\varphi\Psi\right)\delta\varphi' \nonumber \\
      +& 2\beta\left[\varphi\Psi' + \varphi\Psi\eta - \Psi^2 - 2\pi G_* A^4 e^{2\Lambda}\left(\tilde{\epsilon} - 3\tilde{p}\right) 
      \left(1+4\beta\varphi^2\right)\right]\delta\varphi 
      = 4\pi G_* \beta A^4 e^{2\Lambda} \varphi
        \left(\delta \tilde{\epsilon} - 3\delta\tilde{p}\right),  \label{eq:dphi}
\end{align}
where the dot and prime denote the partial derivative with respect to $t$ and $r$, respectively, while $\Psi=\varphi'$ and $\eta = \Phi' - \Lambda' + 2/r$. On the other hand, from Eq. (\ref{eq:conservation}), one can get the perturbation equations as
\begin{gather}
  \ddot{W} =  -\frac{1}{\tilde{p} + \tilde{\epsilon}}e^{2\Phi-2\Lambda}\left[\delta\tilde{p}' 
      + (\Phi' + \beta\varphi\Psi)(\delta \tilde{p} + \delta\tilde{\epsilon})\right], \label{eq:W} \\
  \delta\tilde{\epsilon} = -(\tilde{p} + \tilde{\epsilon})W' - \left[\tilde{\epsilon}' 
     + (\tilde{p} + \tilde{\epsilon})\left(\Lambda' + \frac{2}{r} + 3\beta\varphi\Psi\right)\right]W, \label{eq:de}
\end{gather}
where we use the TOV equation, i.e., $\tilde{p}'=-(\tilde{p}+\tilde{\epsilon})(\Phi' + \beta\varphi\Psi)$, to derive the above equations. In addition to the evolution equations, one can show the relation between $\delta\tilde{p}$ and $\delta\tilde{\epsilon}$ as $\delta\tilde{p}=c_s^2\delta\tilde{\epsilon}$, where $c_s$ denotes the sound speed defined as $c_s^2=\partial\tilde{p}/\partial\tilde{\epsilon}$. Consequently, using Eqs. (\ref{eq:W}) and (\ref{eq:de}), one can derive the evolution equation for $W$;
\begin{align}
  -e^{-2\Phi+2\Lambda}\ddot{W} &+ c_s^2W'' + \left[2c_sc_s' - \Phi' - \beta\varphi\Psi + c_s^2\left(\Lambda'
     + \frac{2}{r} + 3\beta \varphi\Psi \right)\right] W' \nonumber \\
     &+ \left[2c_sc_s'\left(\Lambda' + \frac{2}{r} + 3\beta\varphi\Psi\right)
     - \Phi'' - \beta\Psi^2 - \beta\varphi\Psi' + c_s^2 \left(\Lambda'' - \frac{2}{r^2} + 3\beta\Psi^2 + 3\beta\varphi\Psi'\right)\right]W = 0
     \label{eq:W0}
\end{align}

In order to calculate the evolutions of the variables $\delta \varphi$ and $W$, one should impose the appropriate boundary conditions. That is, the scalar gravitational wave, $\delta \varphi$, becomes the only outgoing wave at the spatial infinity, while the perturbation variables should be regular in the vicinity of stellar center. The regularity conditions near the center can be written as $W=W_c r$ and $\delta\varphi=\delta\varphi_c$, where $W_c$ and $\delta\varphi_c$ are some constants. Additionally, the Lagrangian perturbation of pressure should vanish at the stellar surface. This condition leads to the boundary conditions at the stellar surface as 
\begin{gather}
 W' + \left(\Lambda' + \frac{2}{R} + 3\beta\varphi\Psi\right)W=0.
\end{gather}

%%%%%%%%%%%%%%%%%%%%%%%%%%%%%%%%%%%%%%%%%%%%%%%%
\section{Numerical Results}
\label{sec:IV}
%%%%%%%%%%%%%%%%%%%%%%%%%%%%%%%%%%%%%%%%%%%%%%%%

As shown in the previous section, the matter can oscillate independently of the oscillations of scalar field, i.e., the fluid oscillations depend only on the background scalar field. So, we consider the fluid oscillations before examining the scalar gravitational waves induced by the matter oscillations in \S \ref{sec:IVa}, and then we examine the details of the scalar gravitational waves in \S \ref{sec:IVb}.

%%%%%%%%%%%%%%%%%%%%%%%%%%%%%%%%%%%%%%%%%%%%%%%%
\subsection{Fluid Oscillations}
\label{sec:IVa}
%%%%%%%%%%%%%%%%%%%%%%%%%%%%%%%%%%%%%%%%%%%%%%%%

Assuming a harmonic dependence of time, such as $W(t,r)=W(r)e^{i\omega t}$, the evolution equation for $W$ [Eq. (\ref{eq:W0})] can be reduced to 
\begin{align}
 c_s^2W'' &+ \left[2c_sc_s' - \Phi' - \beta\varphi\Psi + c_s^2\left(\Lambda'
     + \frac{2}{r} + 3\beta \varphi\Psi \right)\right] W' \nonumber \\
     &+ \left[\omega^2e^{-2\Phi+2\Lambda} + 2c_sc_s'\left(\Lambda' + \frac{2}{r} + 3\beta\varphi\Psi\right)
     - \Phi'' - \beta\Psi^2 - \beta\varphi\Psi' + c_s^2 \left(\Lambda'' - \frac{2}{r^2} + 3\beta\Psi^2 + 3\beta\varphi\Psi'\right)\right]W = 0.
\end{align}
With the boundary conditions at the stellar center and surface together with the normalization condition, the problem to solve becomes the eigenvalue problem with respect to the eigenfrequency $\omega$. In particular, we adopt $W_c=1$ at the stellar center as the normalization condition. We remark that in general one can accurately determine the eigenfrequencies with the mode analysis adopted here, compared with the time domain analysis.

\begin{figure*}
\begin{center}
\begin{tabular}{cc}
\includegraphics[scale=0.5]{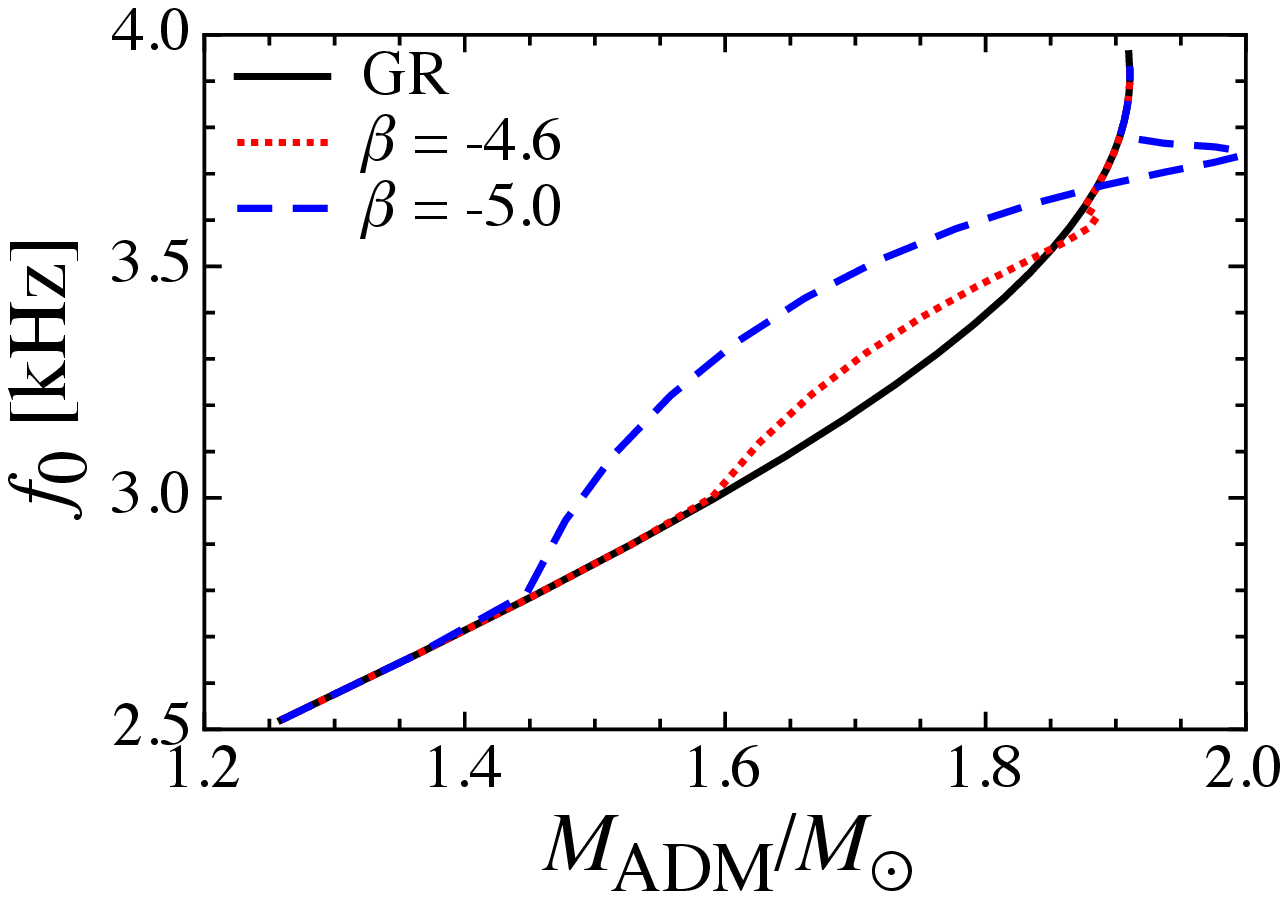} &
\includegraphics[scale=0.5]{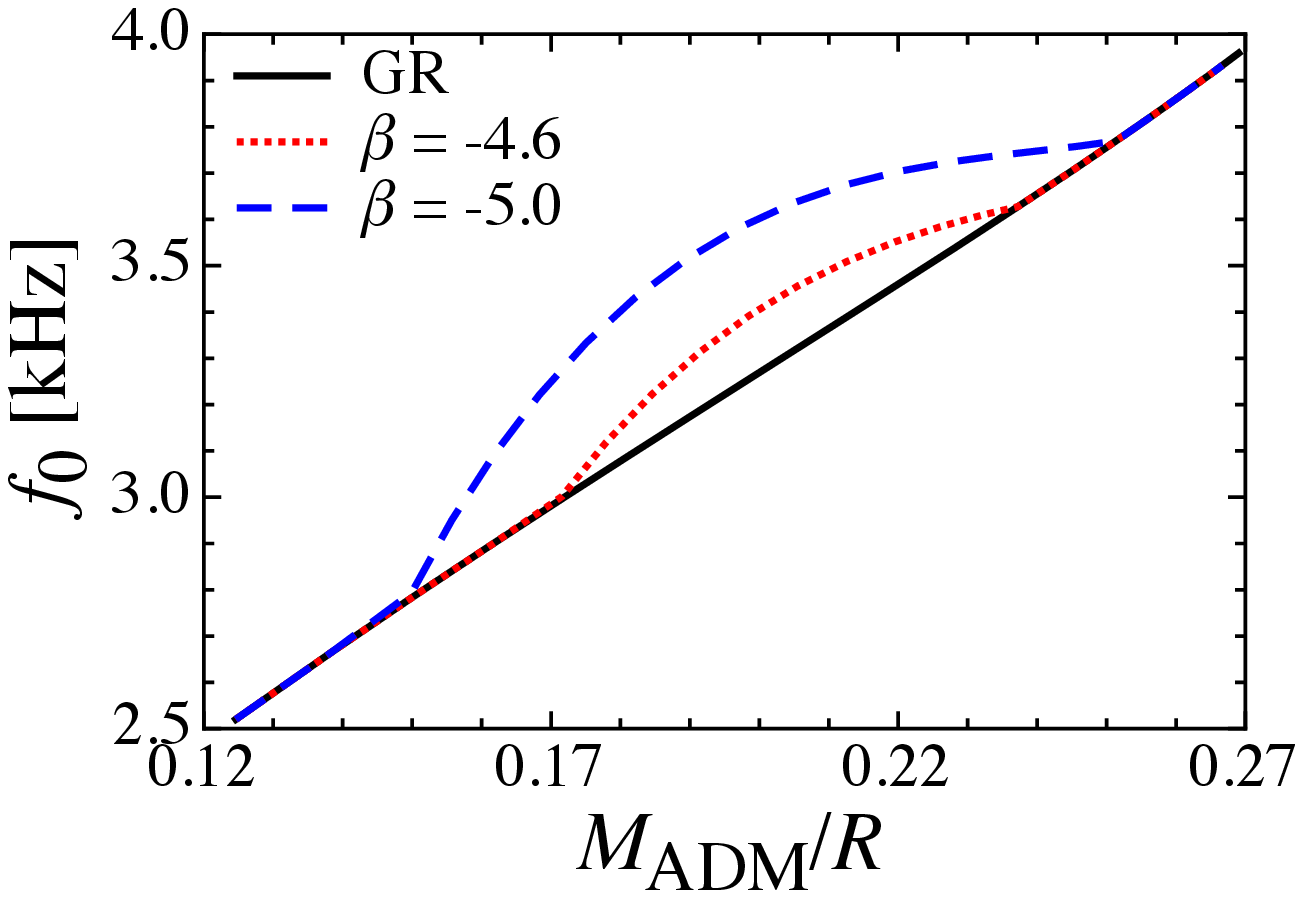}
\end{tabular}
\end{center}
\caption{%%
(Color online) Frequencies of fundamental radial oscillations in scalar-tensor gravity with $\beta=-5.0$ (broken lines) and $-4.6$ (dotted lines), where the left and right panels are functions of ADM mass and stellar compactness, respectively. In addition to the frequencies in scalar-tensor gravity, the results in general relativity are also shown with solid lines. 
}%%
\label{fig:F0}
\end{figure*}

\begin{figure*}
\begin{center}
\begin{tabular}{cc}
\includegraphics[scale=0.5]{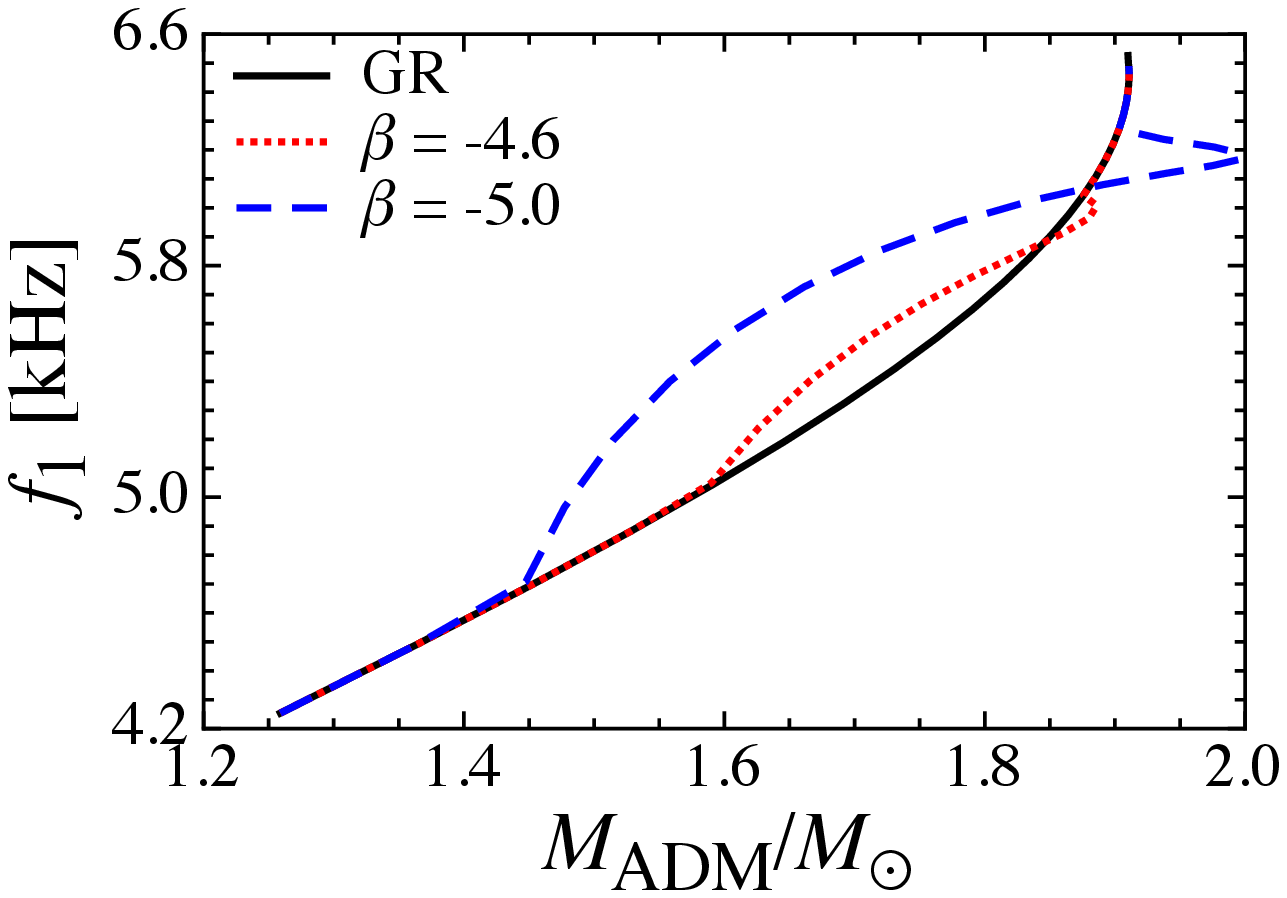} &
\includegraphics[scale=0.5]{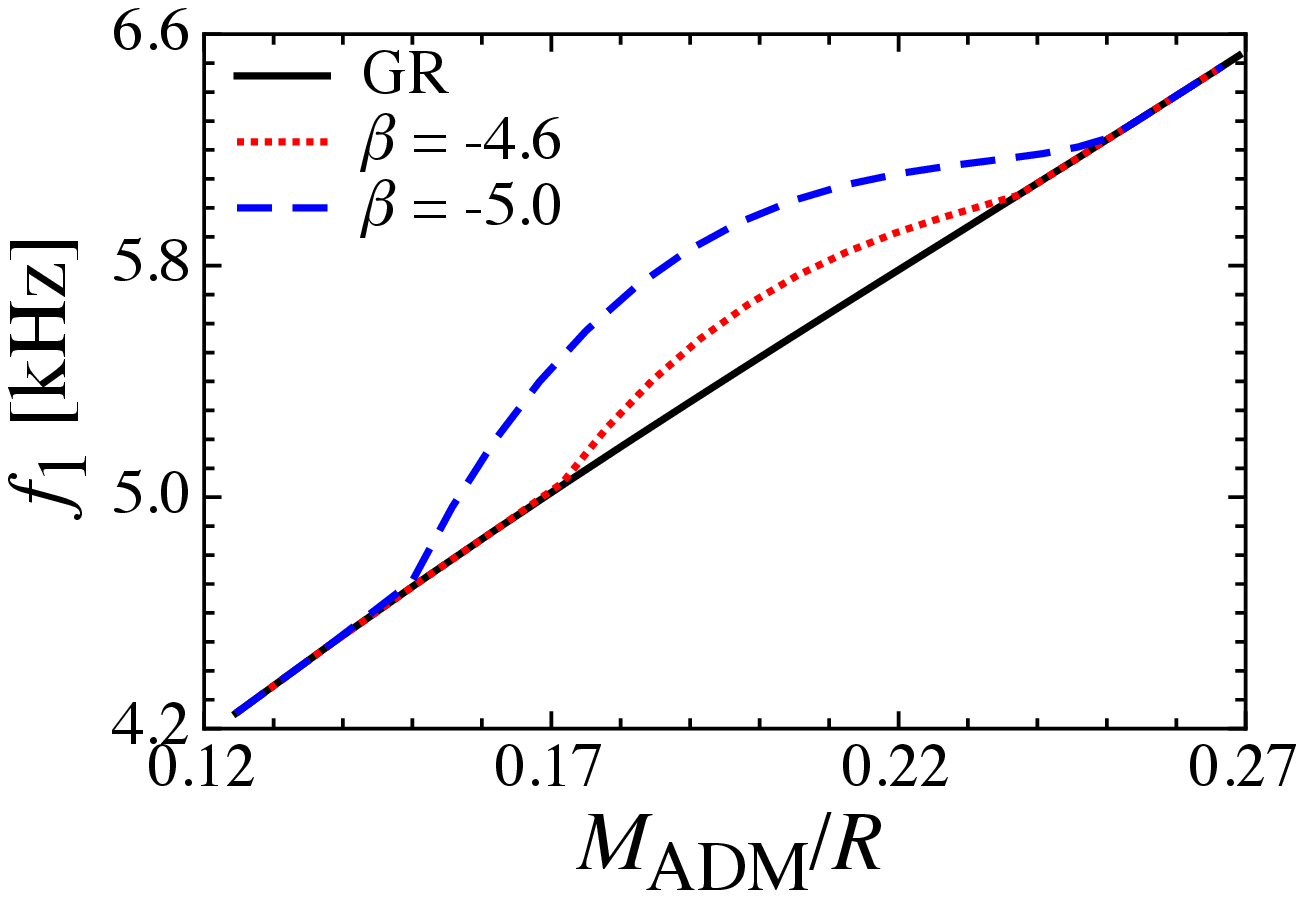}
\end{tabular}
\end{center}
\caption{%%
(Color online) Same as in Fig. \ref{fig:F0}, but for the frequencies of 1st overtone radial oscillations. 
}%%
\label{fig:F1}
\end{figure*}

In Figs. \ref{fig:F0} and \ref{fig:F1}, the frequencies of fundamental and 1st overtone radial oscillations are shown as functions of the ADM mass and stellar compactness in scalar-tensor gravity with $\beta=-5.0$ (broken lines) and $-4.6$ (dotted lines) together with the results in general relativity (solid lines). In these figures, the frequencies $f_i$ are determined as $f_i=\omega_i/2\pi$, where $i$ denotes the nodal number of eigenoscillation. From these figures, one can recognize that the frequencies of radial oscillations in scalar-tensor gravity are affected by the existence of background scalar field, and one can obviously see the spontaneous scalarization even in the frequencies of radial oscillations. That is, one might be able to distinguish the gravitational theory via the observation of radial oscillations with the help of the other observations of the ADM mass or stellar compactness. We remark that the stellar compactness could be determined by the observation of the gravitational redshift of absorption line radiated from the stellar surface.
In practice, if the radial oscillations would be excited in the magnetized neutron stars, the hydromagnetic waves could be also generated, which might become the source of the radiations of electromagnetic waves \cite{C1965,D1989}. If so, one might be possible to detect the imprint of radial oscillations with the electromagnetic waves, which could enable us to distinguish the gravitational theory.

%%%%%%%%%%%%%%%%%%%%%%%%%%%%%%%%%%%%%%%%%%%%%%%%
%  FIGURE 3
%%%%%%%%%%%%%%%%%%%%%%%%%%%%%%%%%%%%%%%%%%%%%%%%
%
%

%%%%%%%%%%%%%%%%%%%%%%%%%%%%%%%%%%%%%%%%%%%%%%%%
%  FIGURE 4
%%%%%%%%%%%%%%%%%%%%%%%%%%%%%%%%%%%%%%%%%%%%%%%%
%
%

%%%%%%%%%%%%%%%%%%%%%%%%%%%%%%%%%%%%%%%%%%%%%%%%
\subsection{Scalar Gravitational Waves}
\label{sec:IVb}
%%%%%%%%%%%%%%%%%%%%%%%%%%%%%%%%%%%%%%%%%%%%%%%%

\begin{figure*}
\begin{center}
\begin{tabular}{cc}
\includegraphics[scale=0.5]{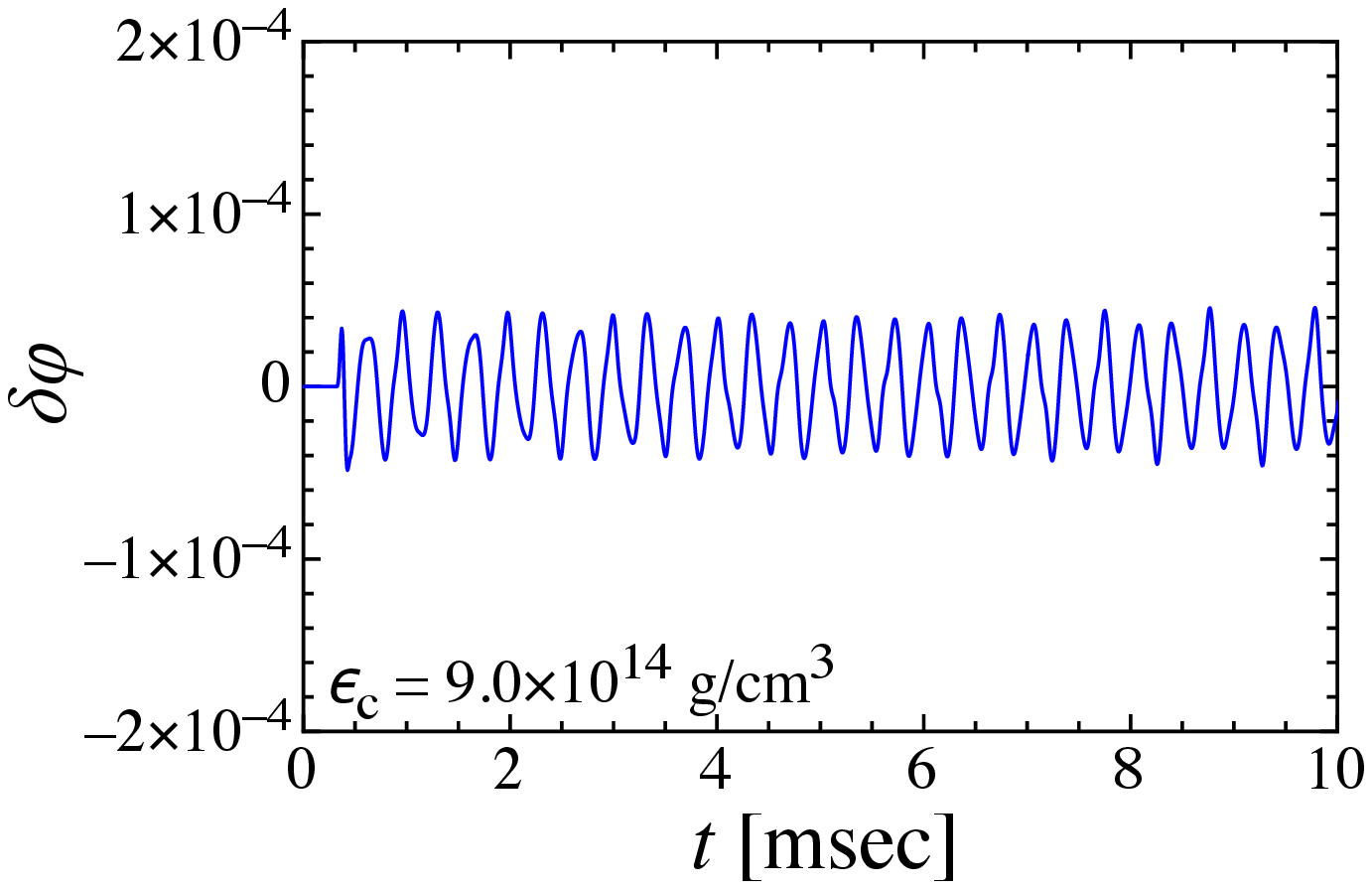} &
\includegraphics[scale=0.5]{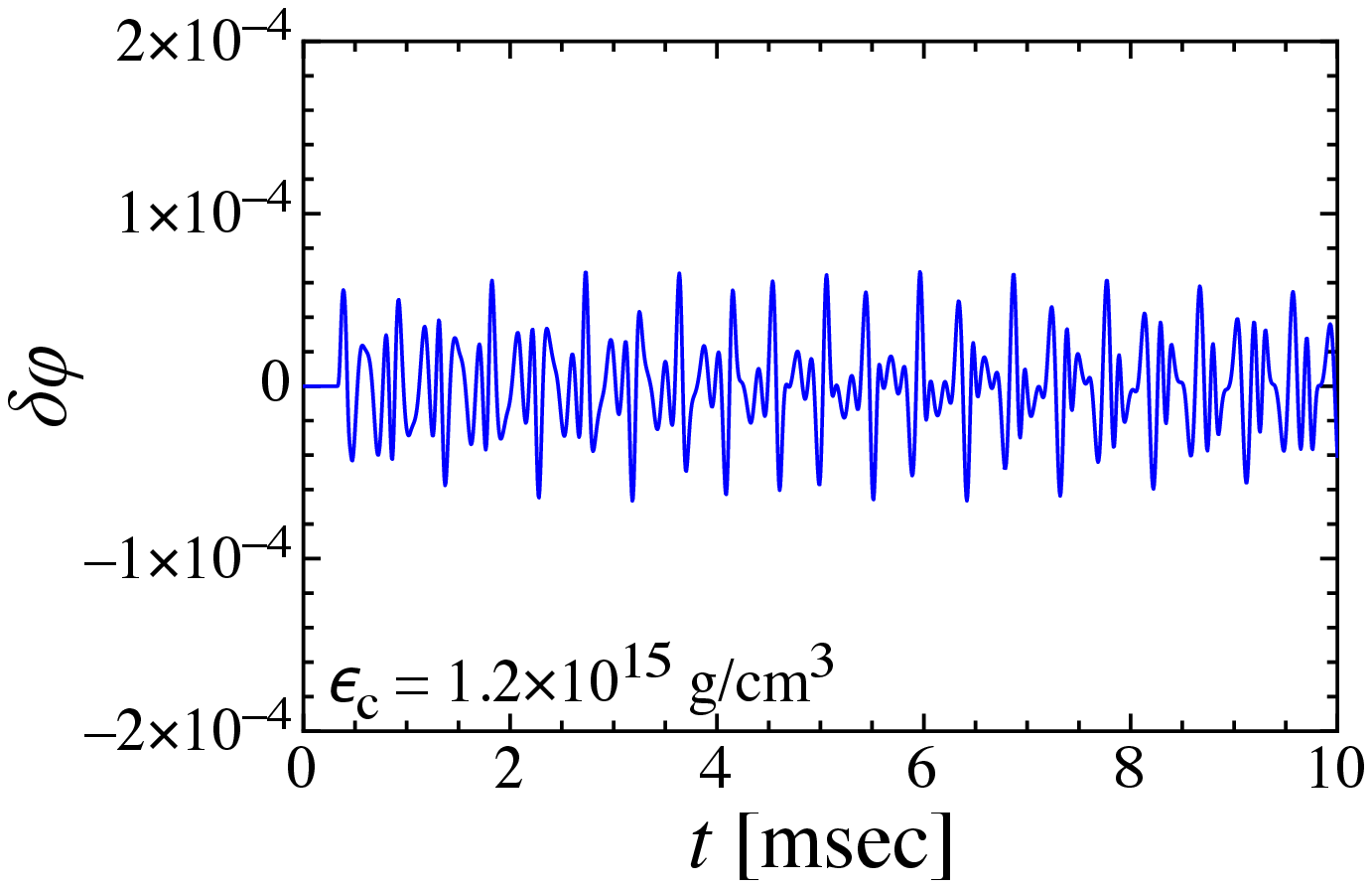} \\
\includegraphics[scale=0.5]{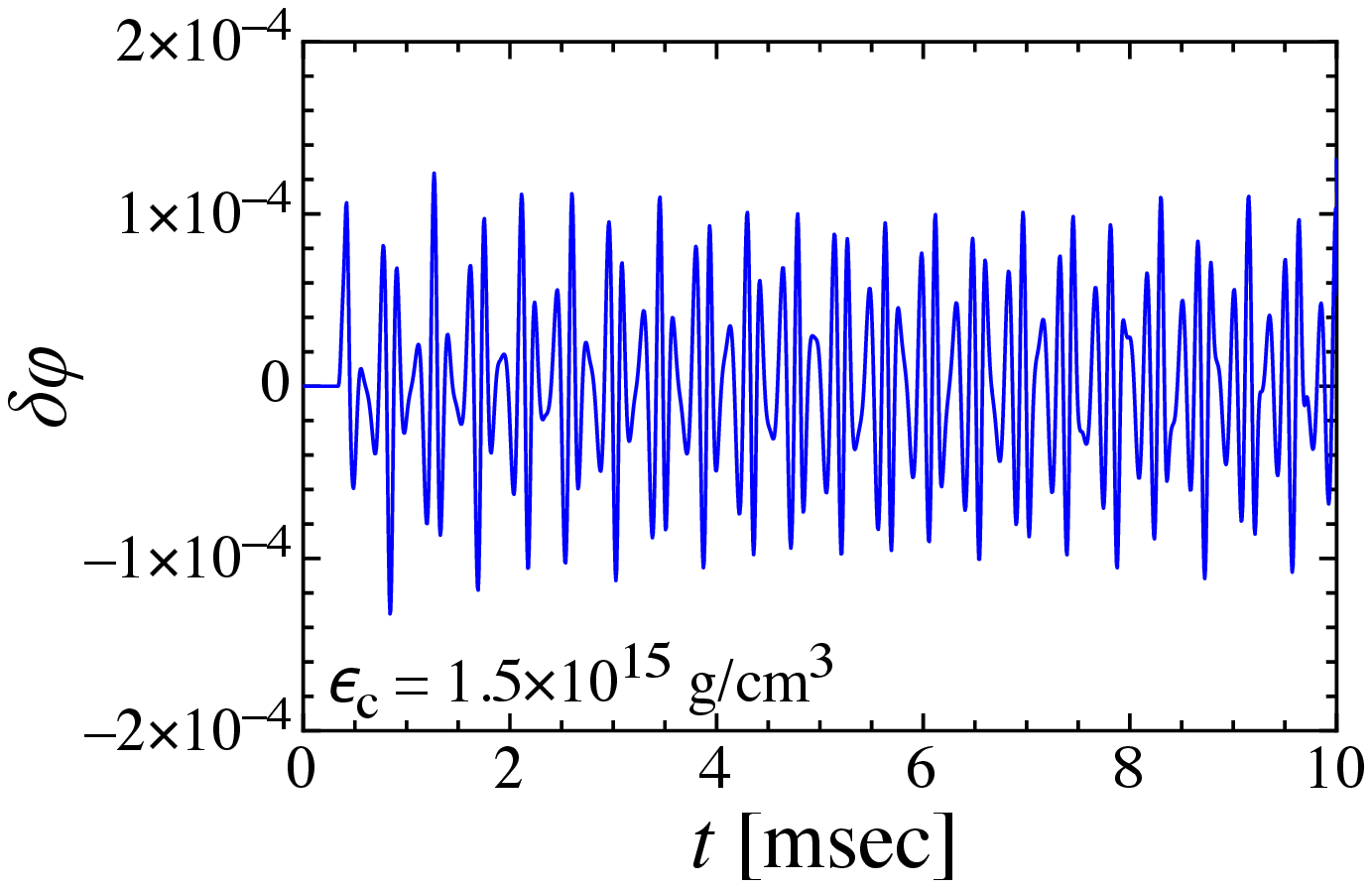} &
\includegraphics[scale=0.5]{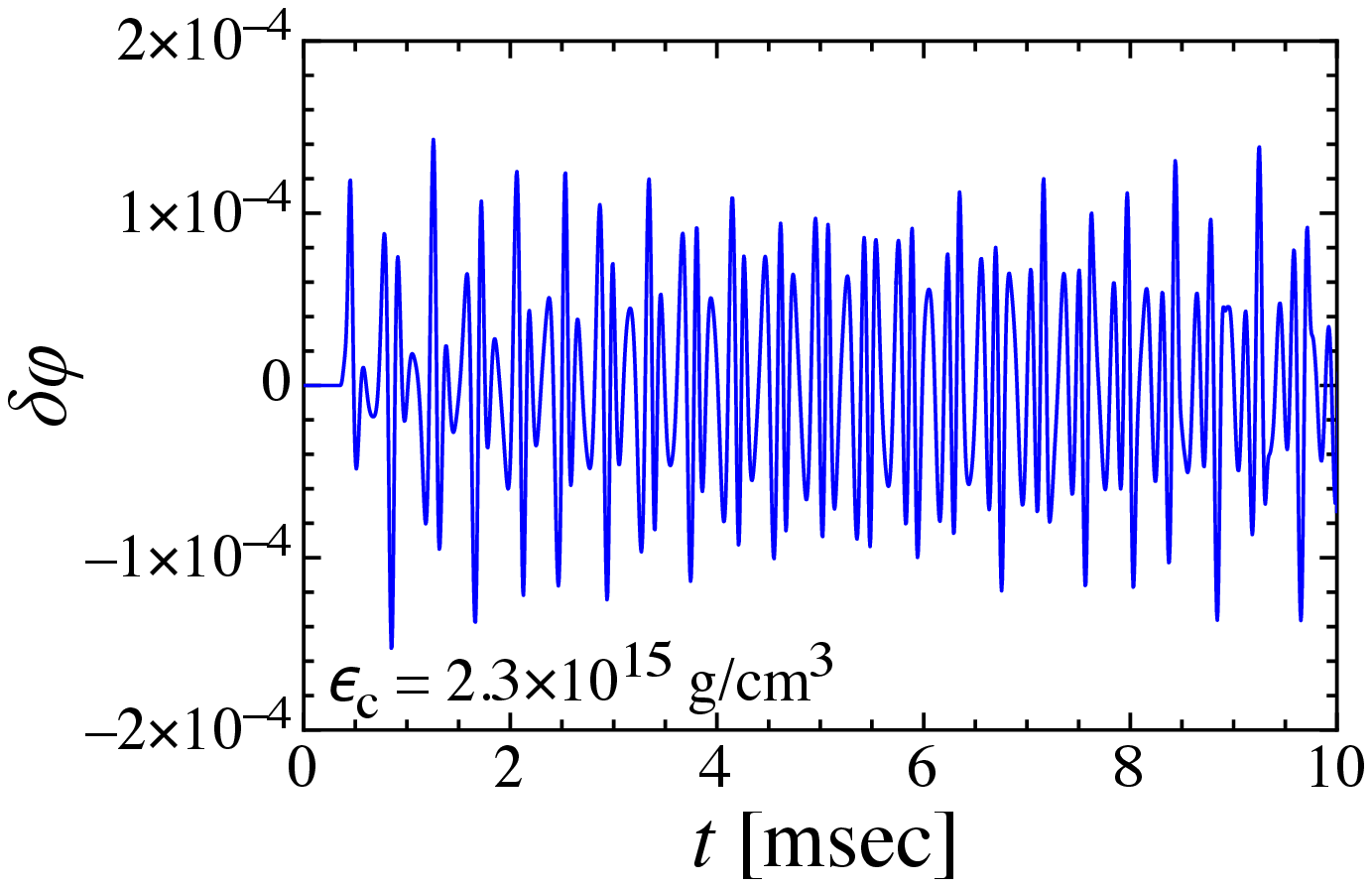} 
\end{tabular}
\end{center}
\caption{%%
(Color online) Waveforms of scalar gravitational waves for $\beta=-5.0$ emitted from the neutron stars constructed with $\epsilon_c=9.0\times 10^{14}$ g/cm$^3$ (upper left panel), $1.2\times 10^{15}$ g/cm$^3$ (upper right panel), $1.5\times 10^{15}$ g/cm$^3$ (lower left panel), and $2.3\times 10^{15}$ g/cm$^3$ (lower right panel).
}%%
\label{fig:WF}
\end{figure*}

Now, we examine the scalar gravitational waves radiated from the neutron stars by calculating the time evolution of Eq. (\ref{eq:dphi}) directly. To systematically examine the scalar gravitational waves induced by the matter oscillations, we consider the zero scalar gravitational wave initially, and put the initial distribution of matter displacement $W_0(r)$ given by
\begin{equation}
 W_0(r) = w\left(\frac{r}{R}\right)\left(\frac{r-R}{R}\right)^2,
\end{equation}
where $w$ is a constant. And then, we determine the value of $w$ in such a way that the initial energy due to the matter oscillations $E_0$ is fixed to especially $E_0=10^{-4}M_\odot$, where $E_0$ is defined as
\begin{equation}
 E_0 = \frac{1}{2}\int\frac{\delta\tilde{p}}{\tilde{\epsilon}}\delta\tilde{\epsilon}^*d^3x.
\end{equation}
The waveforms of scalar gravitational waves driven by the matter oscillations can be calculated as shown in Fig. \ref{fig:WF} for $\beta=-5.0$, where the upper left, upper right, lower left, and lower right panels correspond to the waveforms radiated from the neutron stars constructed with $\epsilon_c=9.0\times 10^{14}$, $1.2\times 10^{15}$, $1.5\times 10^{15}$, and $2.3\times 10^{15}$ g/cm$^3$, respectively. From these panels, one can see that the scalar gravitational waves are excited if the background scalar field is non-zero, where the amplitudes of scalar gravitational waves could depend on the strength of background scalar field.

%%%%%%%%%%%%%%%%%%%%%%%%%%%%%%%%%%%%%%%%%%%%%%%%
%  FIGURE 5
%%%%%%%%%%%%%%%%%%%%%%%%%%%%%%%%%%%%%%%%%%%%%%%%

\begin{figure*}
\begin{center}
\begin{tabular}{cc}
\includegraphics[scale=0.5]{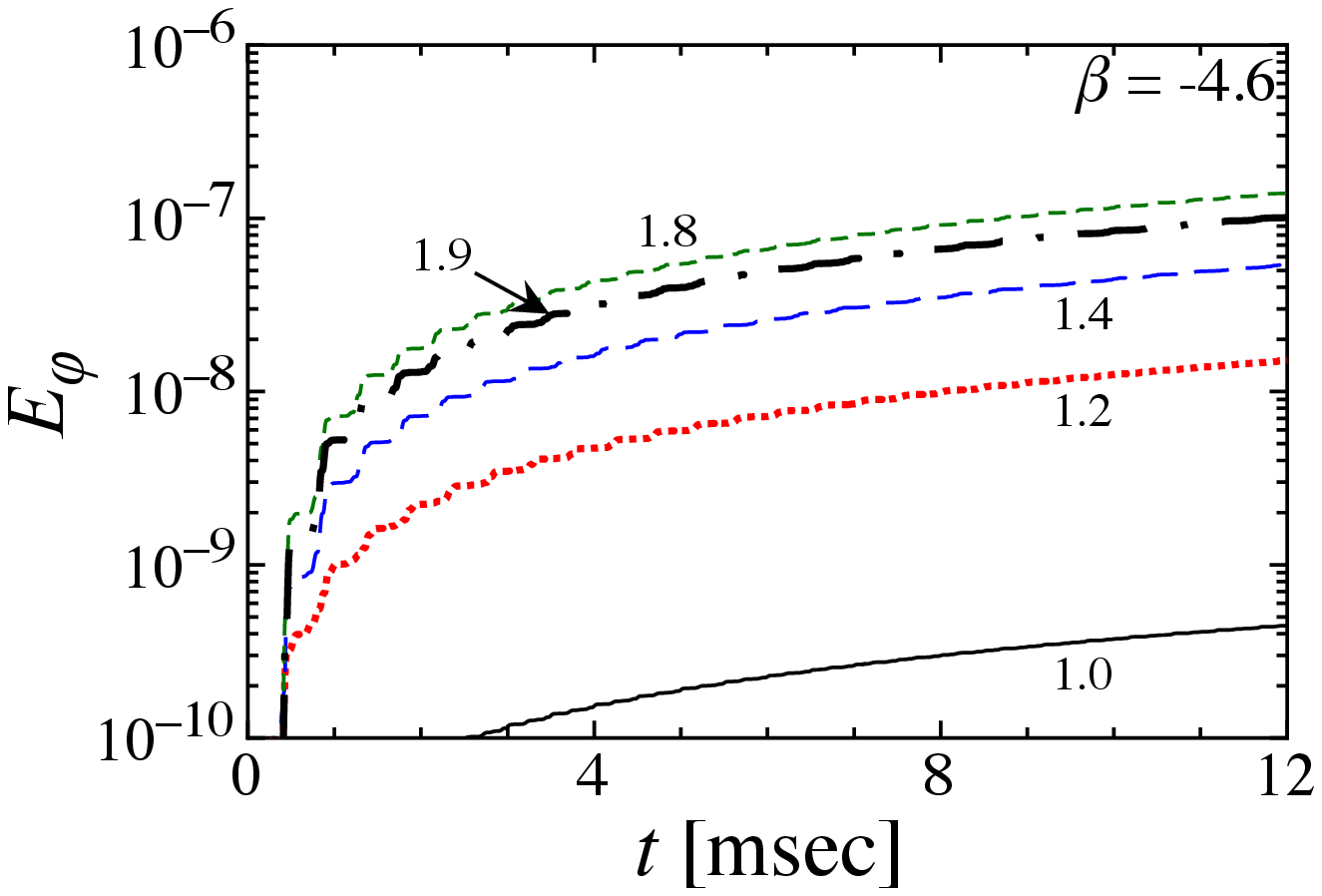} &
\includegraphics[scale=0.5]{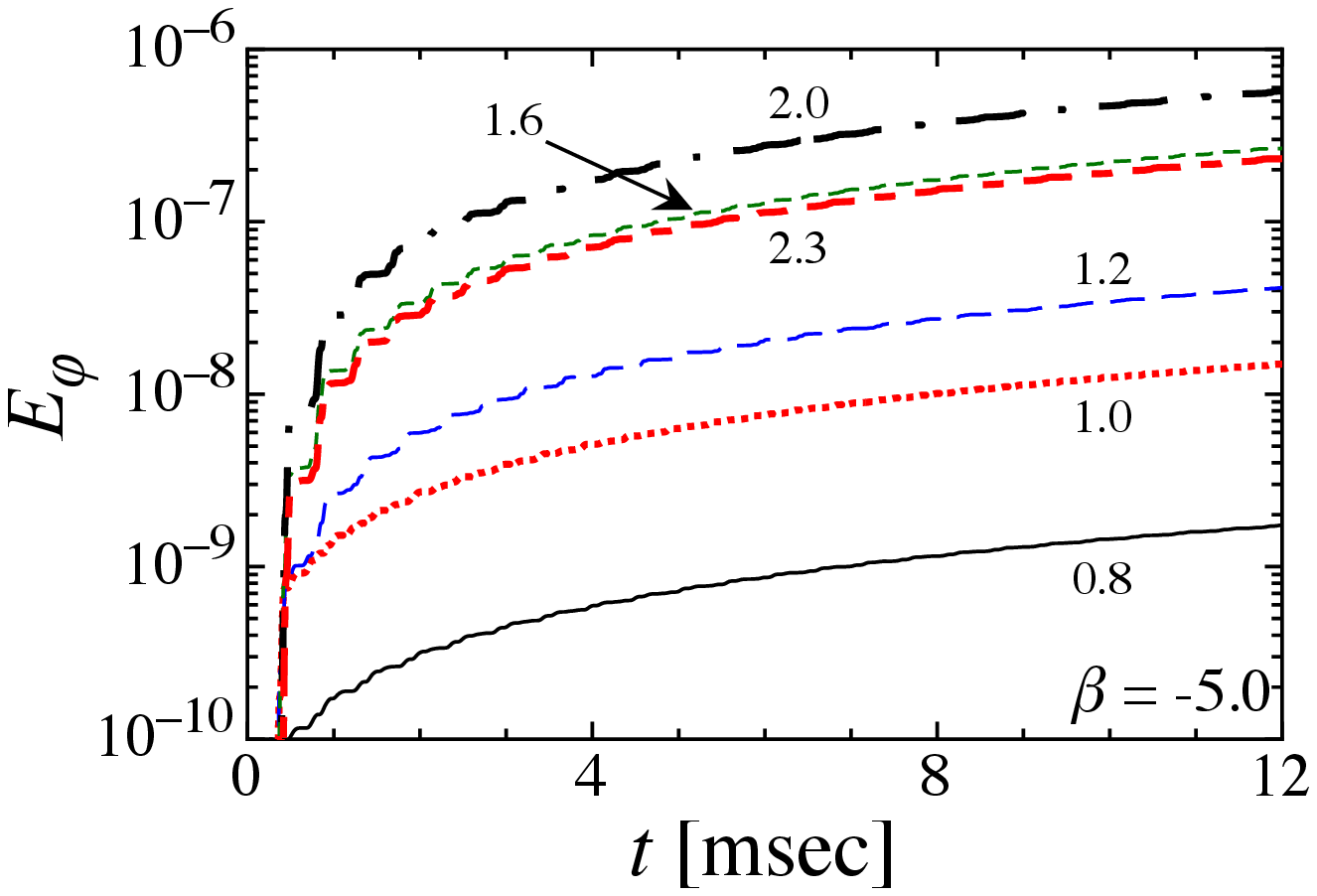}
\end{tabular}
\end{center}
\caption{%%
(Color online) Time evolutions of radiated energy of scalar gravitational waves for the different stellar models with $\beta=-4.6$ (left panel) and $-5.0$ (right panel), where the adopted central densities are shown on each line in the unit of $10^{15}$ g/cm$^3$.
}%%
\label{fig:Ephi}
\end{figure*}

\begin{figure*}
\begin{center}
\includegraphics[scale=0.5]{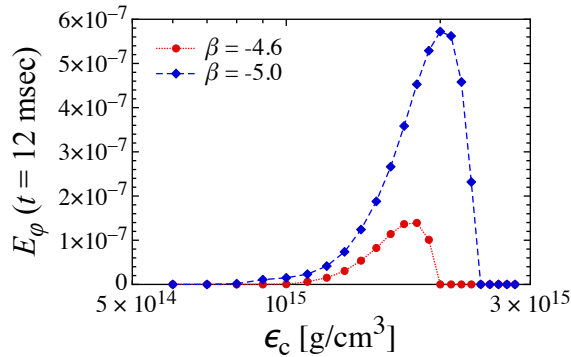}
\end{center}
\caption{%%
(Color online) The total energy radiated by scalar gravitational waves until $t=12$ msec as a function of the stellar central density, where the dotted and broken lines correspond to the results for $\beta=-4.6$ and $-5.0$, respectively.
}%%
\label{fig:Ephi-t12}
\end{figure*}

From observational point of view, the total radiated energy of scalar gravitational waves is an important property, which can be estimated as
\begin{equation}
  E_\varphi(t) \approx \int_0^t |\partial_t \delta\varphi|^2dt.
\end{equation}
Using the numerical data in the evolution of Eq. (\ref{eq:dphi}), the total energies radiated from the different stellar models in scalar-tensor gravity are calculated as shown in Fig. \ref{fig:Ephi}, where the left and right panels correspond to the time evolutions of the total radiated energies for $\beta=-4.6$ and $-5.0$. In each panel, the central densities of the adopted stellar models denote on each line in the unit of $10^{15}$ g/cm$^3$. From this figure, we find that the total energy radiated by the scalar gravitational waves depend strongly on the stellar models. To clearly see the dependence on stellar models, in Fig. \ref{fig:Ephi-t12}, we show the total energies accumulated until $t=12$ msec as a function of the stellar central density for $\beta=-4.6$ (left panel) and $-5.0$ (right panel). Comparing this figure to Fig. \ref{fig:phic}, we can find that the central density for the peak of the total radiated energy is shifted to density region higher than that for the peak of the background scalar field. This means that the massive neutron stars might have a potential to radiate more scalar gravitational waves. Additionally, one observes that the total energy also strongly depends on the coupling parameter $\beta$. In practice, the ratio of the total radiated energy for $\beta=-5.0$ to that for $\beta=-4.6$ reaches $4.1$, while the ratio of the central value of background scalar field for $\beta=-5.0$ to that for $\beta =-4.6$ is only $1.5$.

%%%%%%%%%%%%%%%%%%%%%%%%%%%%%%%%%%%%%%%%%%%%%%%%
%  FIGURE 6
%%%%%%%%%%%%%%%%%%%%%%%%%%%%%%%%%%%%%%%%%%%%%%%%
%
%

%%%%%%%%%%%%%%%%%%%%%%%%%%%%%%%%%%%%%%%%%%%%%%%%
%  FIGURE 7
%%%%%%%%%%%%%%%%%%%%%%%%%%%%%%%%%%%%%%%%%%%%%%%%

\begin{figure*}
\begin{center}
\begin{tabular}{cc}
\includegraphics[scale=0.5]{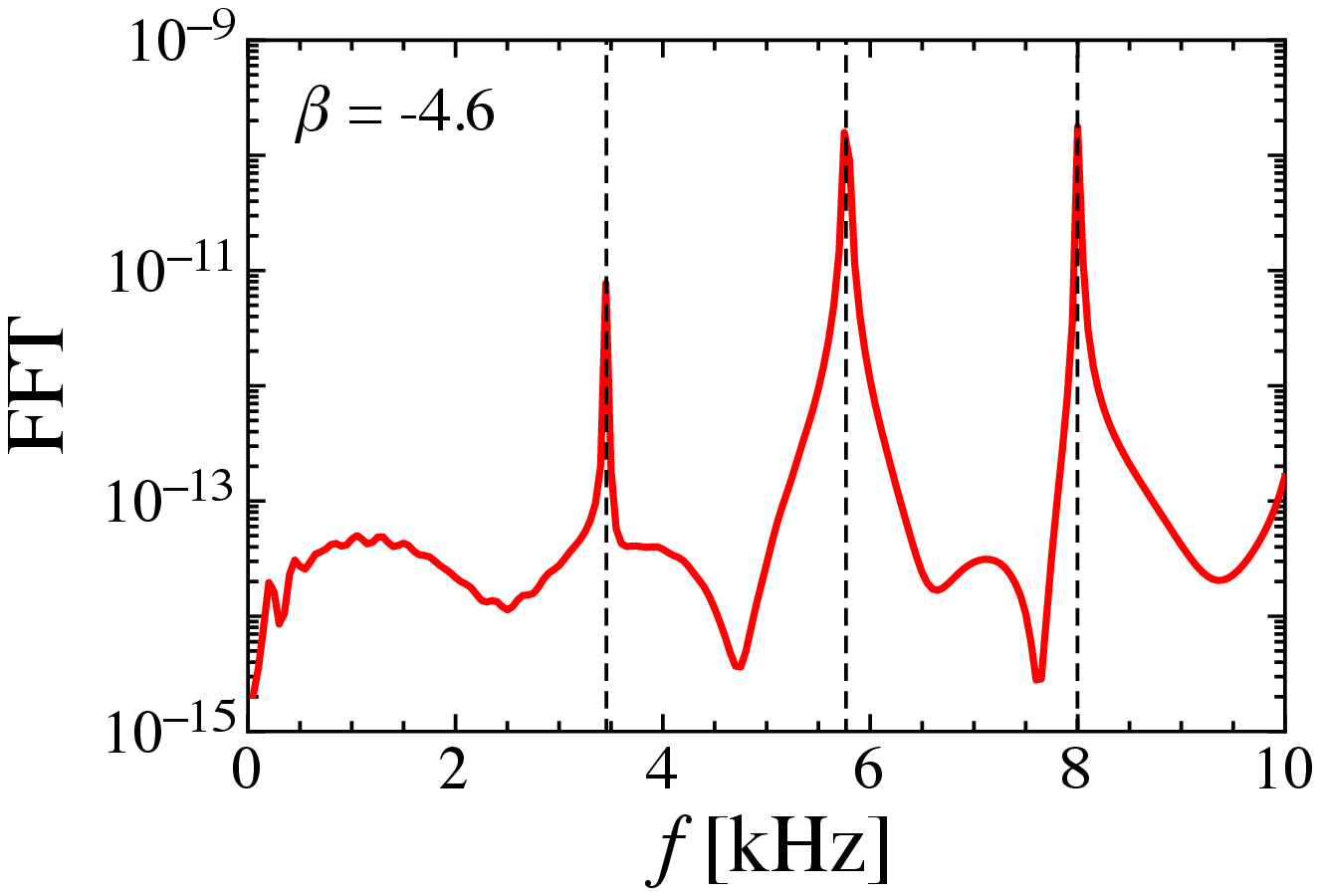} &
\includegraphics[scale=0.5]{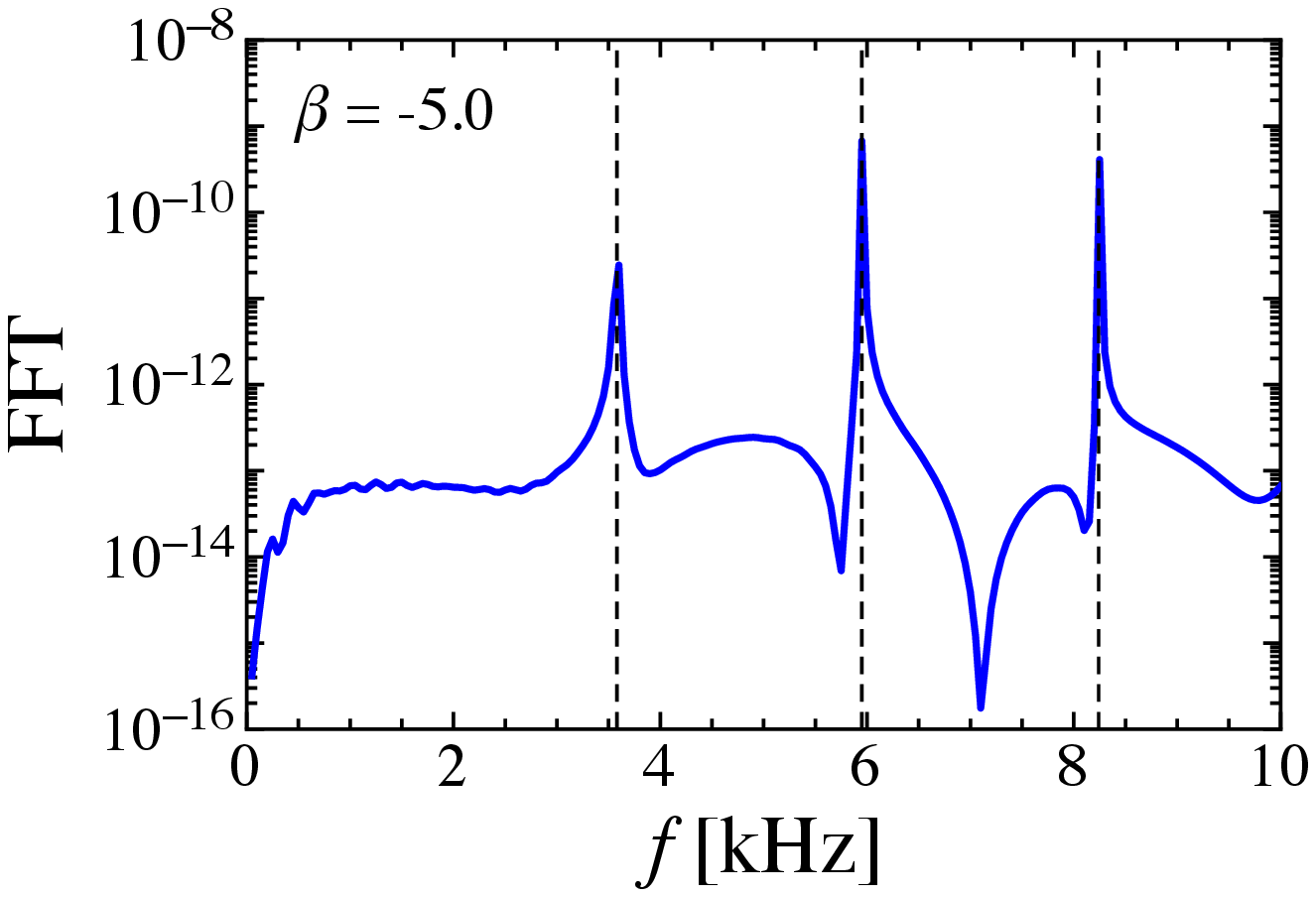}
\end{tabular}
\end{center}
\caption{%%
(Color online) FFT calculated from the waveforms of scalar gravitational waves radiated from the neutron stars with $\epsilon_c=1.5\times 10^{15}$ g/cm$^3$ in scalar-tensor gravity with $\beta=-4.6$ (left panel) and $-5.0$ (right panel). In both panels, the vertical broken lines denote the frequencies of matter oscillations calculated with the eigenvalue problem as in \S \ref{sec:IVa}.
}%%
\label{fig:FFT}
\end{figure*}

Furthermore, in order to see the specific oscillation frequencies of scalar gravitational waves, we calculate the fast Fourier transform (FFT) for the stellar models with $\epsilon_c=1.5\times 10^{15}$ g/cm$^3$ and show it in Fig. \ref{fig:FFT}, where the left and right panels correspond to the results for $\beta=-4.6$ and $-5.0$, respectively. In both panels, we also denote the eigenfrequencies of matter radial oscillations calculated with the mode analysis shown in \S \ref{sec:IVa} with the broken vertical lines. From this figure, one can find that the scalar gravitational waves driven by the matter radial oscillations could oscillate with the same frequencies as those of the matter oscillations. That is, in scalar-tensor gravity, one has a chance to extract the frequencies of radial oscillations of neutron stars via the observations of scalar gravitational waves, which can be written as functions of the stellar mass and/or stellar compactness as in Figs. \ref{fig:F0} and \ref{fig:F1}. This is an advantage in scalar-tensor gravity, because it is impossible to observe the radial oscillations of neutron stars via the radiated gravitational waves in general relativity, where the gravitational waves can not be excited due to the radial oscillations of neutron stars.

%%%%%%%%%%%%%%%%%%%%%%%%%%%%%%%%%%%%%%%%%%%%%%%%
%  FIGURE 8
%%%%%%%%%%%%%%%%%%%%%%%%%%%%%%%%%%%%%%%%%%%%%%%%
%
%

%%%%%%%%%%%%%%%%%%%%%%%%%%%%%%%%%%%%%%%%%%%%%%%%
\section{Conclusion}
\label{sec:V}
%%%%%%%%%%%%%%%%%%%%%%%%%%%%%%%%%%%%%%%%%%%%%%%%

Neutron stars are one of the best candidates to probe the gravitational theory in the strong-field regime. In this paper, we especially focus on the radial oscillations of neutron stars in scalar-tensor gravity to examine the scalar gravitational waves driven by the matter oscillations. In fact, the gravitational waves are not excited due to the radial stellar oscillations in general relativity, while one can expect to observe the scalar gravitational waves due to such oscillations in scalar-tensor gravity. For the calculations of the scalar gravitational waves, we first derive the perturbation equations for radial oscillations in scalar-tensor gravity. From the equation of system of radial oscillations, we find that the matter oscillations depend only on the background scalar field, independently of the scalar gravitational waves. On the other hand, the wave equation of scalar gravitational waves has a source term composed of the matter oscillations. Due to such a specific coupling, we can determine the frequencies of matter radial oscillations by the mode analysis. As a result, we show that the spontaneous scalarization can be observed even in the radial oscillations, which might enable us to find the imprint of gravitational theory with the help of the other observations such as stellar mass and/or compactness.

Additionally, to examine the scalar gravitational waves driven by the matter radial oscillations, we directly make a numerical simulation of the evolution equations, where we fix the initial energy of matter oscillations to be $10^{-4}M_\odot$.  Then, we find that the scalar gravitational waves can be excited if the background scalar field exists. We also find that the total energy radiated by the scalar gravitational waves depends strongly on the background scalar field and the coupling constant $\beta$, where the massive star has a potential to radiate more energy of scalar gravitational waves. Furthermore, we make the fast Fourier transform to see the specific oscillation frequencies of radiated scalar gravitational waves, which are exactly same as the frequencies of matter oscillations. That is, via the observations of scalar gravitational waves, one can extract the frequencies of stellar radial oscillations. This is an advantage in scalar-tensor gravity, because the radial gravitational waves can not be excited in general relativity as mentioned before.
We remark that one might have another chance to observe an imprint of radial oscillation in the gravitational waves, if radial oscillations are strongly excited, for example in core-collapse supernovae, and those oscillations make nonlinear coupling with nonradial oscillations, where the oscillations with combination frequencies could be excited \cite{PSN2007}. If so, one might be possible to measure the background scalar field via such nonlinear coupling.
In this paper, as a first step, we neglect the effects of the solid crust layer, magnetic fields, and the exotic matter inside the star, which are also important properties of neutron stars. Such effects could bring us the additional information about the stellar properties \cite{SKLS2013,Sotani2007,Sotani2011,Yasutake}, which might make observational constraints in the gravitational theory stronger.

%%%%%%%%%%%%%%%%%%%%%%%%%%%%%%%%%%%%%%%%%%%%%%%%
\acknowledgments
%%%%%%%%%%%%%%%%%%%%%%%%%%%%%%%%%%%%%%%%%%%%%%%%

We are grateful to K. D. Kokkotas, D. D. Doneva, and S. S. Yazadjiev for valuable comments.
This work was supported by Grants-in-Aid for Scientific Research on Innovative Areas through No.\ 24105001, and No.\ 24105008 provided by MEXT, by Grant-in-Aid for Young Scientists (B) through No.\ 24740177 provided by JSPS, by the Yukawa International Program for Quark-hadron Sciences, and by the Grant-in-Aid for the global COE program ``The Next Generation of Physics, Spun from Universality and Emergence" from MEXT.

%\appendix
%%%%%%%%%%%%%%%%%%%%%%%%%%%%%%%%%%%%%%%%%%%%%%%%
%\section{Functional forms in the linearized energy-momentum conservation law}   % Appendix C
%\label{sec:appendix_1}
%%%%%%%%%%%%%%%%%%%%%%%%%%%%%%%%%%%%%%%%%%%%%%%%

%%%%%%%%%%%%%%%%%%%%%%%%%%%%%%%%%%%%%%%%%%%%%%%%

\end{document}